# Isospin effects on the system size dependence of balance energy in heavy-ion collisions*

Sakshi Gautam[1], and Aman D. Sood[2]

[1] Department of Physics, Panjab University, Chandigarh, India.
[2] SUBATECH, Laboratoire de Physique Subatomique et des Technologies Associees, Universite de Nantes-IN2P3/CNRS-EMN $ rue Alfred Kastler, F-44072 Nantes, France.

The investigation of system size effects in various phenomena of heavy-ion collisions has attracted a lot of attention [1]. With the availability of high intensity radioactive beams at many facilities, effects of isospin degree of freedom in nuclear reactions can be studied. We aim to study isospin effects on the mass dependence of balance energy using isopsin-dependent quantum molecular dynamics (IQMD) model [2].

To see isospin effects, we take two sets of isobars with N/Z =1 and 1.4 throughout the mass range between 48 and 270. Interestingly, from figure (a) we see that throughout the mass range more neutron rich system has higher balance energy [3]. The calculated balance energies fall on a line that is a power law fit (proportional to $A^\tau$) where $\tau$ = -0.45 and 0.50 for N/Z = 1.4 and 1, respectively. The different values of $\tau$ for two sets can be attributed to the increasing role of Coulomb repulsion in the case of N/Z = 1. To see the role of Coulomb potential, we also show calculations with reduced Coulomb potential. Figure (b) displays the percentage difference between the balance energies for systems having N/Z = 1 and 1.4. We see that percentage difference increases with system size pointing towards the role of Coulomb potential.

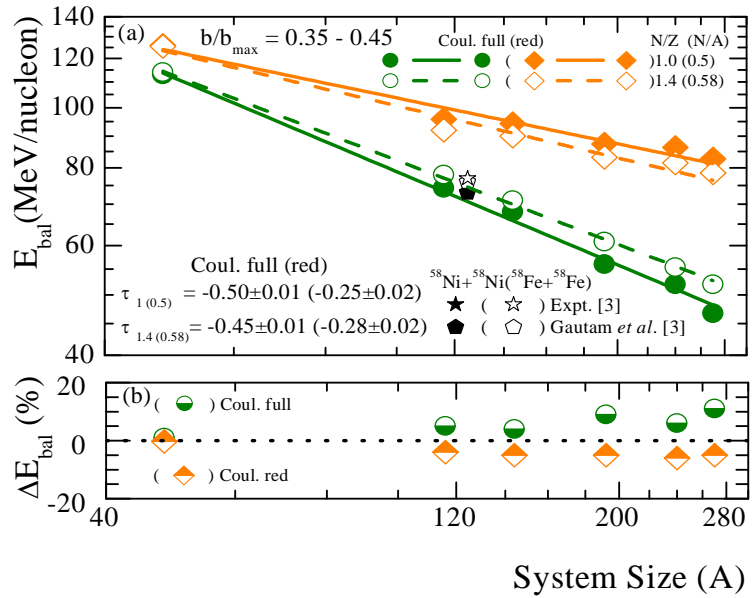

Figure 1: (a) Balance energy as a function of combined mass of the system.(b) The percentage difference as a function of combined mass of the system.

* This work is supported by Indo-French project no. 4104-1.

[1] R. K. Puri and R. K. Gupta, J. Phys G: Nucl. Part. Phys. **18**, 903 (1992).
[2] C. Hartnack *et al* Eur. Phys. J A **1**, 151 (1998).
[3] S. Gautam *et al.*, J. Phys G: Nucl. Part. Phys. (in press); R. Pak *et al.*, Phys. Rev. Lett. **78**, 1022 (1996); ibid. **78**, 1026 (1996); C. Liewen *et al.*, Phys. Rev. C **58**, 2283 (1998).